\documentclass[10pt,preprint]{aastex}

\begin{document}

\shortauthors{Luhman et al.}
\shorttitle{FLAMINGOS Spectroscopy in IC~348}

\title{FLAMINGOS Spectroscopy of New Low-Mass Members of the Young Cluster 
IC~348}

\author{K. L. Luhman\altaffilmark{1}, Elizabeth A. Lada\altaffilmark{2,3}, 
August A. Muench\altaffilmark{1,3}, and Richard J. Elston\altaffilmark{2,3}}

\altaffiltext{1}{Harvard-Smithsonian Center for Astrophysics, 60 Garden St.,
Cambridge, MA 02138, USA; kluhman@cfa.harvard.edu.}

\altaffiltext{2}{Department of Astronomy, The University of Florida,
Gainesville, FL 32611, USA.}

\altaffiltext{3}{
Visiting Astronomer, Kitt Peak National Observatory, National Optical Astronomy 
Observatory, which is operated by the Association of Universities for Research 
in Astronomy, Inc., under cooperative agreement with the National Science 
Foundation.}

\begin{abstract}

We present spectroscopy of candidate stellar and substellar members of
the young cluster IC~348. 
Using the Florida Multi-Object Imaging Near-Infrared Grism Observational 
Spectrometer with the 4 meter telescope at Kitt Peak National Observatory, 
we have obtained multi-object moderate-resolution ($R=1000$) $J$- and $H$-band 
spectra of 66 infrared sources ($H=12$-17) toward IC~348, many of which are 
difficult to observe spectroscopically at optical wavelengths ($I>20$)
because they are highly reddened and/or intrinsically cool and red.
We have also observed 19 known cluster members that have optical spectral 
types available from previous work. 
By using these latter sources as the spectral classification standards, 
we have identified 14 new members of the cluster with types of M2-M6 in the 
sample of 66 new objects. Two additional objects exhibit types
of $>$M8.5, but cannot be conclusively classified as either 
field dwarfs or cluster members with available data.
We have estimated extinctions, luminosities, and effective temperatures for
these 16 M-type objects, placed them on the H-R diagram, and used 
the evolutionary models of Chabrier \& Baraffe to estimate their masses.
If the two candidates at $>$M8.5 are indeed members, they should be among
the least massive known brown dwarfs in IC~348 ($M/M_\odot\sim0.01$).

\end{abstract}


\section{Introduction}
\label{sec:intro}

Young stellar clusters embedded within molecular clouds are valuable 
laboratories for studying the birth of stars, brown dwarfs, and planets 
\citep{ll03}.
Because of its rich and compact nature ($\sim400$~members, $D\sim20\arcmin$) 
and its proximity to the Sun ($d=315$~pc), the IC~348 cluster in the Perseus 
molecular cloud is a particularly appealing population.
Over the last decade, the identification and characterization of the contents 
of young clusters like IC~348 have benefited from advancements in 
detectors, instruments, and telescopes. 
The area and sensitivity of near-infrared (IR) imaging surveys of IC~348 have
progressively increased \citep{ll95,luh98,mue03,pre03}, resulting in 
improved constraints on the mass distribution of cluster members derived 
from statistical analysis of luminosity functions.
A similar trend in optical CCD imaging of IC~348 has extended the 
identification of individual candidate members to
lower masses, higher reddenings, and larger distances from the cluster center
\citep{her98,luh99,luh03b}.
To assess membership and measure spectral types for these candidates, 
increasingly sophisticated spectrometers have been employed, consisting of
optical \citep{her98} and near-IR \citep{luh98} single-slit instruments
and optical multi-slit devices \citep{luh99,luh03b}.

Because of their small angular sizes and the fact that their obscured members 
are brightest beyond 1~\micron, 
embedded clusters are well-suited for multi-object 
spectroscopy at near-IR wavelengths.  
One of the first instruments of this kind is the Florida Multi-Object 
Imaging Near-IR Grism Observational Spectrometer (FLAMINGOS, \citet{el98}),
which we have recently used to expand our spectroscopic work in IC~348, 
particularly among the members at low masses and high reddenings that
are difficult to reach with optical spectroscopy. 
In this paper, we describe the selection of our spectroscopic targets 
and the FLAMINGOS observations (\S~\ref{sec:obs}),
measure the spectral types of the targets (\S~\ref{sec:spt}), 
assess their membership in IC~348 (\S~\ref{sec:mem}), 
estimate the extinctions, effective temperatures, and bolometric luminosities
for 14 confirmed members and 2 late-type candidate members 
and interpret their positions on the Hertzsprung-Russell (H-R) diagram 
with theoretical evolutionary models (\S~\ref{sec:hr}), and discuss
the implications of this work (\S~\ref{sec:disc}). 

\section{Observations}
\label{sec:obs}

For the FLAMINGOS spectroscopy in this study, we selected candidate members 
of IC~348 appearing in the optical and IR color-magnitude diagrams from
\citet{luh03b}. We also included known cluster members that 
have optical spectral types from \citet{luh99} and \citet{luh03b}, which will
be used as the standards during the classification of the candidates in 
\S~\ref{sec:spt}.
After designing a slit mask for a set of these objects, we included
additional slitlets as space allowed for faint IR sources from the images
of \citet{mue03}. In this way, useful spectra were obtained for 19 known
members and 66 previously unclassified sources.

The spectrograph was operated with the 4~m telescope at Kitt Peak National 
Observatory on the nights of 2003 January 15-17 and December 10 and 13.
Spectra were obtained through a different slit mask on each of the five nights.
For each mask, the number of exposures, integration time per exposure, 
and targets are listed in Table~\ref{tab:log}.
The instrument employed a $2048\times2048$ HgCdTe HAWAII-2 array, which 
yielded a plate scale of $0\farcs316$ pixel$^{-1}$ and a field of view of a
$10\farcm8\times10\farcm8$ on the 4~m telescope. 
The width of each mask slitlet was $0\farcs95$ (3 pixels).
This configuration produced full coverage from 1.1 to 1.8~\micron\ with a 
spectral resolution of $R=\lambda/\Delta\lambda\sim1000$. The targets were
dithered between two positions separated by $4\arcsec$ along the slitlets.
A nearby G dwarf was also observed through a long slit for the correction 
of telluric absorption. 
After dark subtraction and flat-fielding, adjacent images
along the slit were subtracted from each other to remove sky emission.  
The sky-subtracted images were aligned and combined.  
A spectrum of each target was then extracted, wavelength calibrated
with OH airglow lines, and divided by the spectrum of the telluric standard.
The intrinsic spectral slope and absorption features of the standard were
removed by multiplying by the solar spectrum. 

\section{Analysis}

\subsection{Spectral Types}
\label{sec:spt}

\begin{figure}
\epsscale{0.7}
\plotone{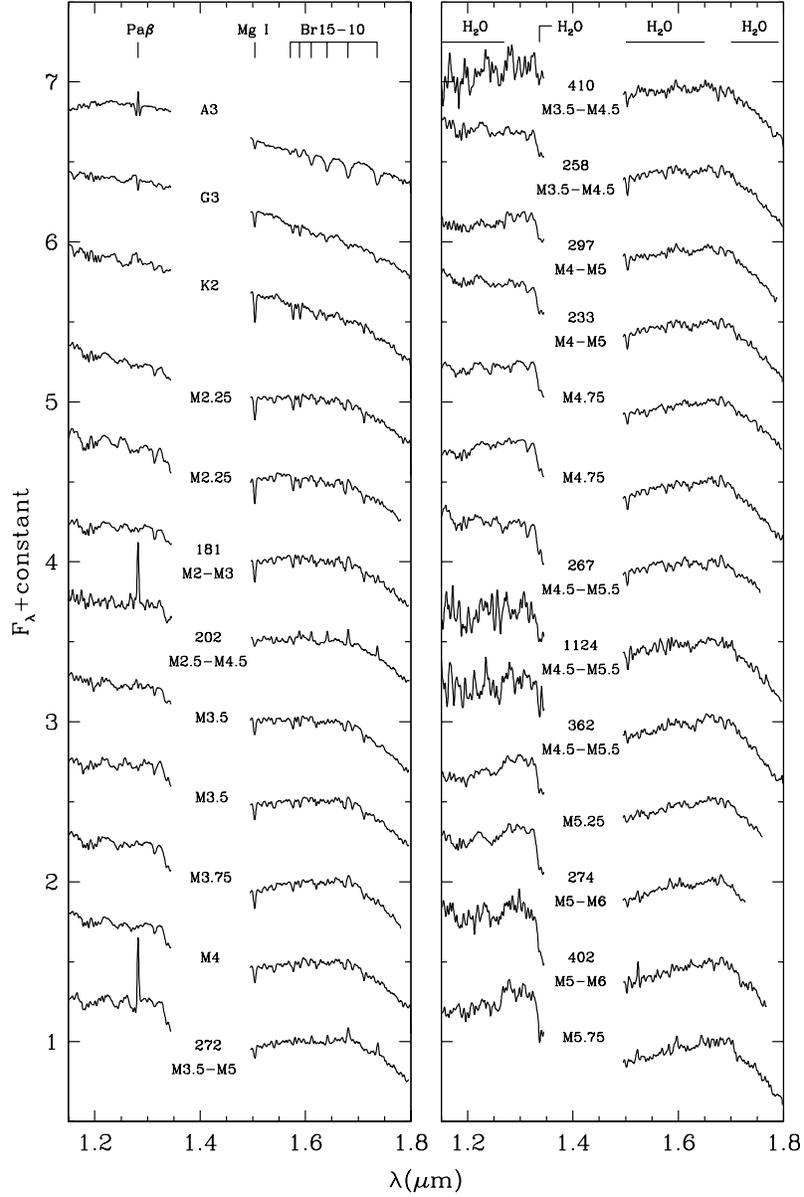}
\caption{
FLAMINGOS near-IR spectra of objects toward the IC~348 cluster with spectral
types earlier than M6. Previously known members are labeled with only their 
optical classifications. 
New members are labeled with their identifications and the spectral types 
derived from a comparison to the IR spectra of the optically-classified members.
The spectra are displayed at a resolution of $R=500$, normalized at 
1.68~\micron, and dereddened (\S~\ref{sec:spt}).
}
\label{fig:spec1}
\end{figure}

\begin{figure}
\plotone{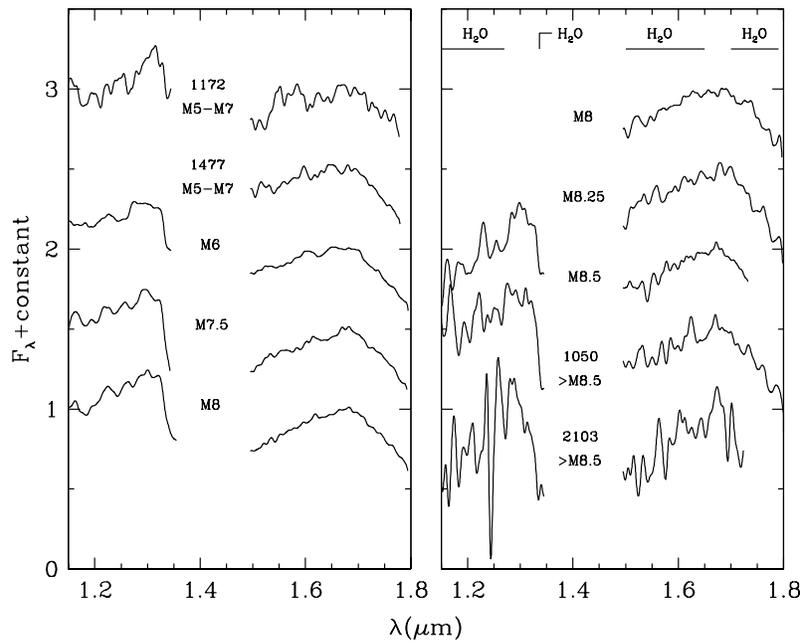}
\caption{
FLAMINGOS near-IR spectra of objects toward the IC~348 cluster with spectral
types of M6 and later. Previously known members are labeled with only their 
optical classifications. New members (1172, 1477) and candidate members 
(1050, 2103) are labeled with their identifications and the spectral types 
derived from a comparison to the IR spectra of the optically-classified members.
The spectra are displayed at a resolution of $R=200$, normalized at 
1.68~\micron, and dereddened (\S~\ref{sec:spt}).
}
\label{fig:spec2}
\end{figure}

To measure spectral types for the candidate members of IC~348 that we
have observed spectroscopically, we begin by examining the spectra of the 
known members with optical classifications in our sample. 
These members consist of sources 12A (A3), 6 (G3), 59 (K2), 120 (M2.25), 122 
(M2.25), 207 (M3.5), 210 (M3.5), 76 (M3.75), 95 (M4), 266 (M4.75), 112 (M4.75), 
230 (M5.25), 555 (M5.75), 298 (M6), 329 (M7.5), 405 (M8), 611 (M8), 
613 (M8.25), and 603 (M8.5), where the identifications and spectral types are
from \citet{luh99} and \citet{luh03b}.
The spectra for the known members are shown in order of optical type in 
Figures~\ref{fig:spec1} and \ref{fig:spec2} and are labeled with only their 
optical classifications. Because the telluric correction was not always 
accurate between 1.345 and 1.495~\micron, data in this wavelength range 
is omitted. The strongest atomic and molecular features in these spectra
are labeled, which consist of the H~I and Mg~I lines and the steam bands.
Other weaker lines are detected as well, identifications for which 
can be found in \citet{wal00}, \citet{mey98}, and \citet{mc04}.
As demonstrated in Figures~\ref{fig:spec1} and \ref{fig:spec2}, the detected IR
spectral features change monotonically with optical type and thus can be
used to classify the new candidates observed in this work. 
The steam bands are particularly useful for measuring spectral types
of late-type objects \citep{wgm99,rei01,leg01,luh04}, as shown by the rapid 
change in the decrement at 1.35~\micron\ and in the broad absorption at both
ends of the $H$-band. Because of the broad nature of these bands,
they can be measured at low spectral resolution. 
Meanwhile, the narrow atomic lines become weak and difficult to detect
at late M types, and thus higher resolution becomes unnecessary.
Therefore, the signal-to-noise of the coolest and faintest objects can
be improved by smoothing to a lower resolution without compromising the 
spectral diagnostics. The spectra for objects later 
than M6 are smoothed to a resolution of $R=200$ in Figure \ref{fig:spec2},
while the earlier objects in Figures~\ref{fig:spec1} are shown at a 
resolution of $R=500$.

Because near-IR steam absorption bands are stronger in young objects
than in field dwarfs at a given optical spectral type \citep{lr99,luc01,mc04},
spectral types of young objects derived from steam with dwarf standards
will be systematically too late. 
Instead, to arrive at accurate spectral types, optically-classified 
young objects rather than dwarfs should be used when measuring spectral types
of young sources from steam \citep{lr99,luh03b}, which is the approach we adopt 
in the following classification of the candidate members of IC~348.

Extinction within the natal cloud of IC~348 results in large
variations in the slopes of the observed spectra in our sample.
To facilitate the comparison of the steam band depths between the candidates and
the optically-classified known members, we have dereddened all spectra with
detectable steam to the same slope as measured by the ratios of fluxes at 
1.32 and 1.68~\micron. These dereddened spectra are not meant to be 
precise estimates of
the intrinsic, unreddened appearance of these stars since the slopes
likely vary with spectral type. After dereddening, for each candidate
we measured a spectral 
type by visually comparing the steam bands and atomic lines in its spectrum
to those in the data of the optically-classified members. Through this analysis,
we find that 16 of the 66 new objects in our sample exhibit M types. 
Within the sequence of optically-classified members in Figures~\ref{fig:spec1} 
and \ref{fig:spec2}, these 16 sources have been inserted and are labeled with 
their identifications and the types derived from these IR spectra.
These objects are also listed Table~\ref{tab:mem}.
The other 50 sources lack steam absorption and therefore are earlier than 
$\sim$M0. More accurate classifications for these early-type stars 
are not possible in most cases because the signal-to-noise is too low to 
measure the atomic lines. However, the spectral type constraints are 
sufficient to indicate that all of these 50 objects without steam are probably 
field stars (\S~\ref{sec:mem}). These stars are listed in Table~\ref{tab:field}.

A few of the new M-type objects require additional comments.
Continuum emission from cool circumstellar material should be much weaker
than photospheric emission in bands as short as $J$ and $H$, particularly 
given that a small fraction of members of IC~348 have noticeable excess 
emission at $K$ \citep{hai01}. However, for two of the new sources, 
202 and 272, the $H$-band atomic lines are weaker than expected for the 
spectral type implied by the steam bands, which is suggestive of modest 
continuum veiling. The presence of veiling is not surprising given
the strong H emission in both stars, which is a signature of active accretion,
and the large $K$-band excess in source 202. Because of this possible veiling,
the uncertainties in the derived spectral types for these stars are larger 
than for the other new M-type objects.
Finally, the steam absorption bands in sources 1050 and 2103 are stronger 
than in the latest optically-classified member in our spectroscopic sample 
(M8.5). Therefore, we can place only upper limits on the spectral types
of these two objects. 

\subsection{Membership}
\label{sec:mem}

A star projected against IC~348 could be a member of the cluster or
a field star in the foreground or the background. For each of the 66
previously unclassified sources in our spectroscopic sample, we employ the 
diagnostics described by \citet{luh03b} to distinguish between these 
possibilities. 

Objects that appear near or below the main sequence when placed on the
H-R diagram at the distance of IC~348 are likely to be background field stars. 
In the previous section, 50 stars were found to lack steam absorption, which
implied spectral types earlier than M0. Some of these stars have even
earlier limits based on the presence of hydrogen absorption lines. 
These spectral type constraints indicate that all of these stars are likely
to be background field stars because of their low positions on the H-R diagram.
In other words, given their faint magnitudes, these sources should be much
cooler than observed if they were members of IC~348. 
As noted in previous studies of IC~348 and other young clusters 
(e.g., \citet{luh03b}), young stars that are seen in scattered light can 
appear low on the H-R diagram, even below the main sequence. Indeed, a few
of the previously known members of IC~348 are subluminous in this manner.
As a result, it is possible that 
some of the objects that we classify as background stars in 
Table~\ref{tab:field} could be scattered-light members of IC~348. 
Imaging of IC~348 with the {\it Spitzer Space Telescope} should easily identify 
any young objects of this kind through detections of their mid-IR excesses.

We now evaluate the membership of the remaining objects, which are the 
16 M-type sources.  Because IC~348 is a star-forming cluster, signatures of 
youth comprise evidence of membership.
Examples include emission in the Brackett and Paschen series of hydrogen and 
excess emission at $K$, which are found in three of our sources. 
Next, the presence of significant reddening in the spectrum or colors ($A_V>1$) 
of a star and a position above the main sequence for the distance of IC~348 
indicate that it cannot be a dwarf in the foreground or the background of 
the cluster, respectively. Fourteen of the new M-type objects are not field
dwarfs when this criterion is applied with the extinctions estimated in the 
next section. The remaining two M-type sources,
the ones classified as $>$M8.5, also appear to exhibit reddening in their
spectra and colors. However, the reddenings implied by the spectra are 
uncertain because of the low signal-to-noise of those data while the
reddenings measured from near-IR colors are uncertain because the spectral 
type, and thus the adopted intrinsic colors, are not well constrained. 
Therefore, with this diagnostic, we cannot rule out the possibility that these
two sources are foreground dwarfs.

Pre-main-sequence objects can also be distinguished from field dwarfs 
and giants with spectral features that are sensitive to surface gravity, 
a variety of which have been identified for cool objects at both optical
\citep{mar96,luh99,mc04} and near-IR \citep{luh98,gor03,mc04} wavelengths.
For instance, the broad plateaus observed in the $H$ and $K$ spectra of 
late-M and L dwarfs \citep{rei01,leg01} are absent in data for young objects,
resulting in sharply peaked, triangular continua \citep{luc01}, as illustrated
by the known late-type members of IC~348 in Figure~\ref{fig:spec2}. 
The $H$-band spectra of the new M-type sources appear to exhibit this 
signature of youth, but we do not consider this definitive evidence of 
membership since we do not have spectra of field dwarfs and giants to compare
to these sources.

\begin{figure}
\epsscale{1}
\plotone{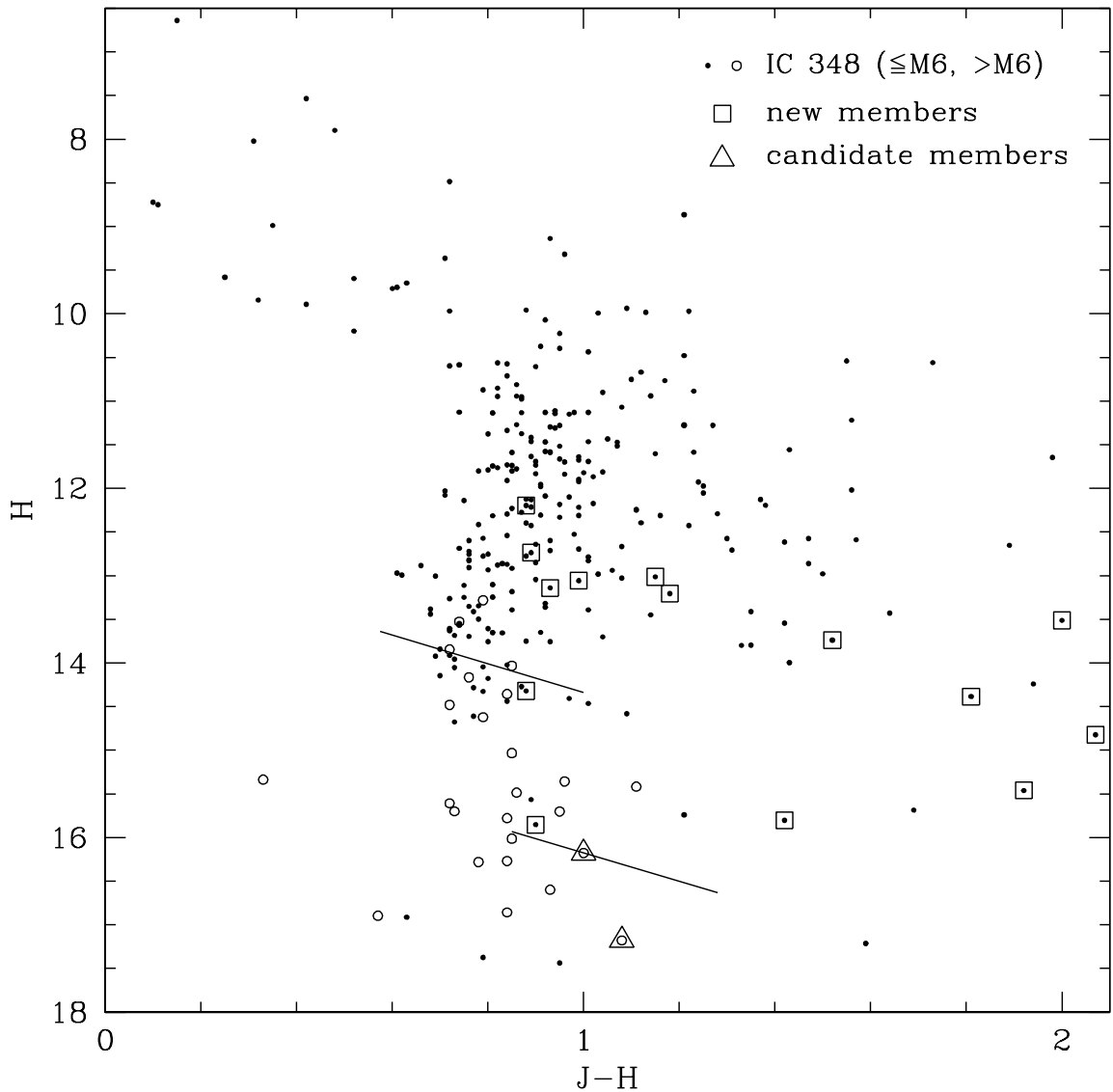}
\caption{
$J-H$ versus $H$ for all known members of the IC~348 cluster at $\leq$M6 and 
$>$M6 ({\it points and circles}) that have been identified through 
spectroscopy in this work ({\it boxes}) and in previous studies.
We also include two late-type objects from our spectroscopic sample 
that lack definitive classifications of membership ({\it triangles}).
The reddening vectors from $A_V=0$-4 are plotted for 0.08 ($\sim$M6.5) and
0.02~$M_\odot$ ($\sim$M9) for ages of 3~Myr ({\it upper and lower solid lines})
\citep{cha00}.
Most of these $J$ and $H$ measurements are from 2MASS and \citet{mue03}.
}
\label{fig:jh}
\end{figure}

The results of our analysis of the membership of the 16 new M-type sources 
are compiled in Table~\ref{tab:mem}. In summary, we have found evidence of 
membership in IC~348 for 14 of the 16 new M-type sources. The remaining 
two objects are promising candidate members that require additional data
to determine their membership definitively.
The masses and reddenings of these 16 sources relative to those of the
previously known members of IC~348 are illustrated by the diagram of $J-H$
versus $H$ in Figure~\ref{fig:jh}.

\subsection{H-R Diagram}
\label{sec:hr}

In this section, we estimate effective temperatures and bolometric
luminosities for the 14 new members of IC~348, place these data on the
H-R diagram, and use theoretical evolutionary models to infer masses and ages.
We also estimate the values of these properties expected for the two late-M 
candidate members, 1050 and 2103, if they are bonafide members.

For each source, we adopt the average extinction measured from its
$I-Z$ and $J-H$ colors. The extinction from $I-Z$ is computed in the manner
described by \citet{luh03b}. For $J-H$, we compute extinction from the color
excess relative to dwarf colors at the spectral type in question, with the
exception of sources 202 and 1124. Because these stars exhibit large
$K$-band excesses, their $J-H$ and $H-K_s$ colors are instead dereddened to 
the locus observed for classical T~Tauri stars \citep{mey97}.
Spectral types are converted to temperatures with the temperature scale from
\citet{luh03b}. Bolometric luminosities are estimated by combining
$J$-band measurements, a distance modulus of 7.5, and bolometric corrections
described by \citet{luh99}. The estimates of extinctions, temperatures,
and luminosities for the 16 M-type sources are listed in Table~\ref{tab:mem}.

\begin{figure}
\plotone{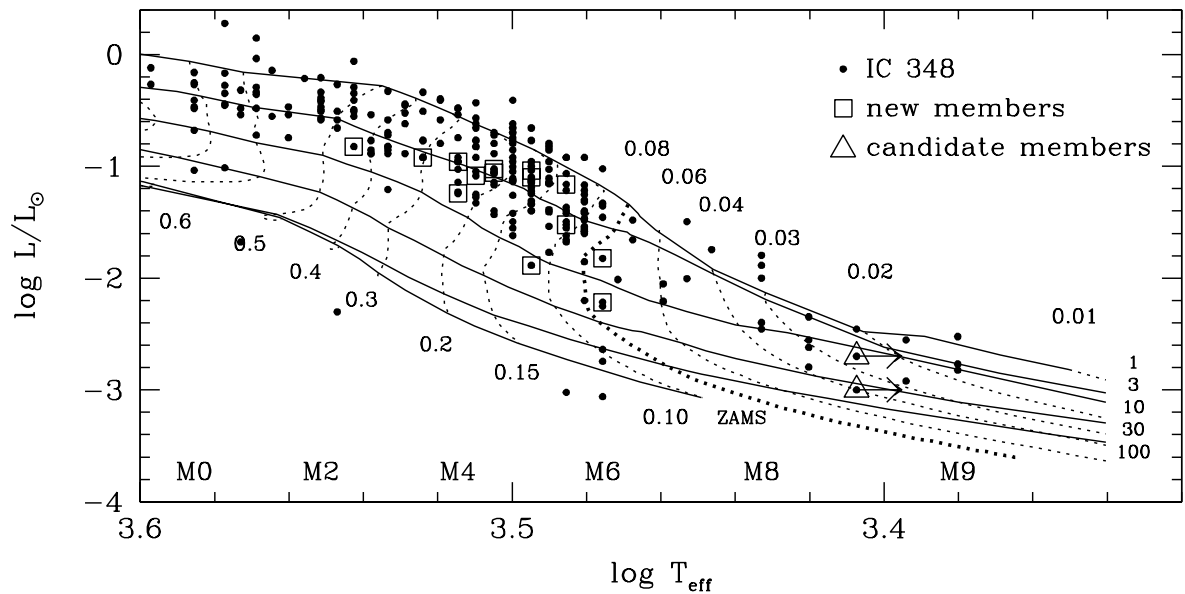}
\caption{
H-R diagram for the IC~348 cluster with the same symbols as in 
Figure~\ref{fig:jh}, except that both early and late-type members are shown
as points. We include the theoretical evolutionary models of
\citet{bar98} ($M/M_\odot>0.1$) and \citet{cha00} ($M/M_\odot\leq0.1$),
where the mass tracks ({\it dotted lines}) and isochrones ({\it solid lines}) 
are labeled in units of $M_\odot$ and Myr, respectively. 
The data for the previously known members are from \citet{luh03b}. 
}
\label{fig:hr}
\end{figure}

These temperatures and luminosities can be interpreted in terms of masses 
and ages with theoretical evolutionary models. 
We select the models of \citet{bar98} and \citet{cha00} because they provide
the best agreement with observational constraints \citep{luh03b}.
The 14 new members of IC~348 and the two candidate members are plotted 
with these models on the H-R diagram in Figure~\ref{fig:hr}. For comparison,
we also include the sequence of previously known members from \citet{luh03b},
which encompasses the positions of the new sources. The 14 new members
at M2-M6 exhibit masses of 0.45 to 0.08~$M_\odot$ according to the adopted
models. Meanwhile, if the two late-type candidates are cluster members,
then they should have masses less than 0.04~$M_\odot$. 
Adopting the median cluster age of $\sim2$~Myr implies masses of only 
$\sim0.01$~$M_\odot$ for these two sources.

\section{Discussion}
\label{sec:disc}

We have demonstrated that multi-object spectroscopy at near-IR
wavelengths can be successfully used to efficiently obtain accurate
spectral classifications ($\pm$0.5-1~subclasss at M types) for
large numbers of faint candidate young stars and brown dwarfs in embedded 
clusters. Our FLAMINGOS spectroscopy of 66 IR sources toward
IC~348 has resulted in the discovery of 14 new members, several of which
are highly reddened and inaccessible at optical wavelengths. 
These sources exhibit spectral types of M2-M6, corresponding to masses
from 0.45~$M_\odot$ to near the hydrogen burning mass limit. 
Two additional sources in our spectroscopic sample exhibit strong 
steam absorption bands that are indicative of types later than M8.5.
If these objects are members of IC~348 rather than field dwarfs, they should
have masses of $\sim0.01$~$M_\odot$, and thus comprise the least massive 
known brown dwarfs in the cluster.  
Accurate spectral types for these two candidates will require either 
optical spectroscopy or IR spectra at higher signal-to-noise than provided 
in this work. Classification of the latter data would require IR spectra of
optically-classified young objects at types later than the limit of M8.5
in our sample. Near-IR spectra at higher signal-to-noise also would 
provide the measurements of gravity-sensitive features and reddening that
are needed to conclusively establish these candidates as either field
dwarfs or young members of IC~348.

\acknowledgements
K. L. was supported by grant NAG5-11627 from the NASA Long-Term Space 
Astrophysics program. We thank Nick Raines for invaluable technical support and
thank Bruno Ferreira, Charles Lada, Joanna Levine, Nick Raines, 
Noah Rashkind, Carlos Roman, Aaron Steinhauer and Andrea Stolte for
mask preparation and assistance with the observations.
The FLAMINGOS data were obtained under the NOAO
Survey Program ``Towards a Complete Near-Infrared Imaging and Spectroscopic
Survey of Giant Molecular Clouds" (PI: E. Lada) and supported by NSF grants
AST97-3367 and AST02-02976 to the University of Florida. FLAMINGOS was
designed and constructed by the IR instrumentation group
(PI: R. Elston) at the Department of Astronomy at the University of Florida
with support from NSF grant AST97-31180 and Kitt Peak National Observatory.

\begin{deluxetable}{lllll}
\tabletypesize{\scriptsize}
\rotate
\tablewidth{0pt}
\tablecaption{Observing Log \label{tab:log}}
\tablehead{
\colhead{} &
\colhead{} &
\colhead{} &
\colhead{$\tau_{exp}$} &
\colhead{} \\
\colhead{Mask} &
\colhead{Date} &
\colhead{$N_{exp}$} &
\colhead{(min)} &
\colhead{ID}}
\startdata
1 & 2003 Jan 15 & 12 & 10 & 12A,59,95,112,120,202,297,333,362,410,469,574,581,584,588,594,901,917,943,951,959,998,1477,1735,2103 \\
2 & 2003 Jan 16 & 12 & 10 & 6,207,210,230,266,298,405,485,546,555,745,4053 \\
3 & 2003 Jan 17 & 4 & 10 & 222,329,392,603,611,613 \\
4 & 2003 Dec 10 & 12 & 5 & 109,112,122,267,356,411,444,466,502,509,1007,1020,1021,1050,1062,1555,1729,1742,3076,3104,3117,3215 \\
5 & 2003 Dec 13 & 8 & 5  & 76,181,233,238,258,272,274,375,377,402,429,448,1110,1124,1139,1142,1167,1172,1173,1194,22253 \\
\enddata
\end{deluxetable}

\begin{deluxetable}{lllllllllllllll}
\tabletypesize{\scriptsize}
\rotate
\tablewidth{0pt}
\tablecaption{New Members and Candidate Members of IC 348 \label{tab:mem}}
\tablehead{
\colhead{ID} &
\colhead{$\alpha$(J2000)\tablenotemark{a}} &
\colhead{$\delta$(J2000)\tablenotemark{a}} &
\colhead{Spectral Type} &
\colhead{Adopt} &
\colhead{Member\tablenotemark{b}} &
\colhead{$T_{\rm eff}$\tablenotemark{c}} &
\colhead{$A_J$} & \colhead{$L_{\rm bol}$} &
\colhead{$R-I$\tablenotemark{d}} & \colhead{$I$\tablenotemark{e}} &
\colhead{$I-Z$\tablenotemark{f}} &
\colhead{$J-H$\tablenotemark{g}} & \colhead{$H-K_s$\tablenotemark{g}} & \colhead{$K_s$\tablenotemark{g}} \\
\colhead{} &
\colhead{h m s} &
\colhead{$\arcdeg$ $\arcmin$ $\arcsec$} &
\colhead{} &
\colhead{} &
\colhead{-ship} &
\colhead{} &
\colhead{} &
\colhead{} &
\colhead{} &
\colhead{} &
\colhead{} &
\colhead{} &
\colhead{} &
\colhead{}
}
\startdata
     181 &   03 44 35.89 &    32 15 53.4 &  M2-M3 &       M2.5 &         $A_V$ &   3488 &    0.60 &    0.15 &      \nodata &    14.90 &     0.64 &     0.88 &     0.24 &    11.96  \\
     202 &   03 44 34.28 &    32 12 40.7 & M2.5-M4.5 &       M3.5 &   $A_V$,e,ex &   3342 &    2.88 &    0.12 &      \nodata &    20.17 &     1.33 &     2.00 &     1.35 &    12.16  \\
     233 &   03 44 41.88 &    32 17 56.7 &  M4-M5 &       M4.5 &         $A_V$ &   3198 &    0.71 &   0.088 &      \nodata &    15.90 &     0.80 &     0.89 &     0.35 &    12.39  \\
     258 &   03 44 43.30 &    32 17 57.1 & M3.5-M4.5 &         M4 &         $A_V$ &   3270 &    1.47 &    0.11 &    2.39 &    17.07 &     0.99 &     1.15 &     0.49 &    12.53  \\
     267 &   03 44 31.83 &    32 15 46.5 & M4.5-M5.5 &         M5 &         $A_V$ &   3125 &    1.53 &   0.092 &    2.36 &    17.47 &     1.11 &     1.18 &     0.50 &    12.71  \\
     272 &   03 44 34.13 &    32 16 35.7 & M3.5-M5 &      M4.25 &      $A_V$,e &   3234 &    1.02 &   0.082 &      \nodata &    16.76 &     0.88 &     0.99 &     0.59 &    12.47  \\
     274 &   03 44 48.84 &    32 18 46.6 &  M5-M6 &       M5.5 &         $A_V$ &   3058 &    0.86 &   0.069 &    2.38 &    16.88 &     0.99 &     0.93 &     0.47 &    12.67  \\
     297 &   03 44 33.21 &    32 12 57.5 &  M4-M5 &       M4.5 &         $A_V$ &   3198 &    2.42 &   0.095 &    3.05 &    19.71 &       \nodata &     1.52 &     0.81 &    12.93  \\
     362 &   03 44 42.30 &    32 12 28.2 & M4.5-M5.5 &         M5 &         $A_V$ &   3125 &    3.19 &   0.080 &      \nodata &    21.09 &     1.59 &     1.81 &     1.01 &    13.38  \\
     402 &   03 44 45.56 &    32 18 20.0 &  M5-M6 &       M5.5 &         $A_V$ &   3058 &    1.09 &   0.030 &    2.49 &    18.37 &     1.16 &     0.88 &     0.55 &    13.77  \\
     410 &   03 44 37.56 &    32 11 55.9 & M3.5-M4.5 &         M4 &         $A_V$ &   3270 &    3.46 &   0.057 &      \nodata &    21.75 &     1.44 &     2.07 &     1.01 &    13.81  \\
    1050 &   03 44 34.90 &    32 15 00.0 &    $>$M8.5 &         \nodata &         \nodata &   $<$2555 &     $\sim0.4$ &   $\sim0.002$ &      \nodata &    20.34 &     1.21 &     1.00 &     0.60 &    15.58  \\
    1124 &   03 44 56.74 &    32 17 03.8 & M4.5-M5.5 &         M5 &      $A_V$,ex &   3125 &    2.43 &   0.013 &    2.82 &    20.31 &     1.26 &     1.92 &     1.36 &    14.10  \\
    1172 &   03 44 58.36 &    32 18 11.8 &  M5-M7 &         M6 &         $A_V$ &   2990 &    0.84 &  0.0061 &      \nodata &    19.83 &     1.08 &     0.90 &     0.64 &    15.21  \\
    1477 &   03 44 36.25 &    32 13 04.6 &  M5-M7 &         M6 &         $A_V$ &   2990 &    2.30 &   0.015 &      \nodata &    21.46 &     1.53 &     1.42 &     0.80 &    15.00  \\
    2103 &   03 44 14.92 &    32 13 43.5 &    $>$M8.5 &         \nodata &         \nodata &   $<$2555 &     $\sim0.5$ &   $\sim0.001$ &      \nodata &    21.94 &     1.21 &     1.08 &     0.78 &    16.40  \\
\enddata
\tablenotetext{a}{From the 2MASS Point Source Catalog for 297 and
from \citet{luh03b} for the remaining objects.}
\tablenotetext{b}{Membership in IC~348 is indicated by $A_V\gtrsim1$ and
a position above the main sequence for the distance of IC~348 (``$A_V$"),
$K$-band excess emission (``ex"), or emission in the Brackett and Paschen 
lines of hydrogen (``e"). Sources 1050 and 2103 
lack definitive evidence of membership from available data.}
\tablenotetext{c}{Temperature scale from \citet{luh03b}.}
\tablenotetext{d}{\citet{luh99}.} 
\tablenotetext{e}{From \citet{luh99} for 297 and from \citet{luh03b} for the 
remaining objects.}
\tablenotetext{f}{\citet{luh03b}.}
\tablenotetext{g}{\citet{mue03}.}
\end{deluxetable}

\begin{deluxetable}{lllllllll}
\tabletypesize{\scriptsize}
\tablewidth{0pt}
\tablecaption{Probable Background Stars\label{tab:field}}
\tablehead{
\colhead{ID} &
\colhead{$\alpha$(J2000)\tablenotemark{a}} &
\colhead{$\delta$(J2000)\tablenotemark{a}} &
\colhead{$R-I$\tablenotemark{b}} & \colhead{$I$\tablenotemark{c}} &
\colhead{$I-Z$\tablenotemark{d}} &
\colhead{$J-H$\tablenotemark{e}} & \colhead{$H-K_s$\tablenotemark{e}}
& \colhead{$K_s$\tablenotemark{e}} \\
\colhead{} &
\colhead{h m s} &
\colhead{$\arcdeg$ $\arcmin$ $\arcsec$} &
\colhead{} &
\colhead{} &
\colhead{} &
\colhead{} &
\colhead{} &
\colhead{}
}
\startdata
     109 &   03 44 40.14 &    32 14 28.1 &      \nodata &    14.92 &     0.79 &     0.93 &     0.40 &    11.27  \\
     222 &   03 44 49.61 &    32 09 12.2 &    1.09 &    15.09 &     0.57 &     0.76 &     0.24 &    12.31  \\
     238 &   03 45 04.66 &    32 16 39.0 &      \nodata &    15.40 &     0.64 &     0.82 &     0.25 &    12.45  \\
     333 &   03 44 27.68 &    32 13 55.3 &    1.59 &    17.38 &     0.76 &     1.20 &     0.35 &    13.48  \\
     356 &   03 45 00.12 &    32 13 24.8 &    1.29 &    16.58 &     0.63 &     0.97 &     0.24 &    13.40  \\
     375 &   03 44 50.66 &    32 17 19.2 &    2.30 &    18.41 &     0.92 &     1.24 &     0.45 &    13.91  \\
     377 &   03 44 38.12 &    32 16 45.2 &    1.51 &    17.67 &     0.70 &     0.92 &     0.31 &    13.87  \\
     392 &   03 44 53.62 &    32 08 58.9 &    1.13 &    16.80 &     0.57 &     0.77 &     0.23 &    13.99  \\
     411 &   03 45 05.66 &    32 14 17.1 &    1.27 &    17.11 &       \nodata &     0.90 &     0.19 &    14.18  \\
     429 &   03 44 51.90 &    32 16 35.5 &    1.68 &    17.86 &     0.81 &     1.05 &     0.35 &    13.99  \\
     444 &   03 44 42.35 &    32 15 13.2 &    1.92 &    19.15 &     0.91 &     1.22 &     0.50 &    14.51  \\
     448 &   03 44 47.61 &    32 17 14.4 &    1.70 &    19.12 &     0.82 &     1.23 &     0.46 &    14.79  \\
     466 &   03 44 30.40 &    32 14 58.2 &    1.54 &    18.45 &     0.80 &     1.05 &     0.34 &    14.78  \\
     469 &   03 44 49.47 &    32 12 18.1 &    1.63 &    18.92 &     0.70 &     1.20 &     0.47 &    14.84  \\
     485 &   03 44 32.38 &    32 08 03.2 &    1.73 &    18.92 &     0.82 &     1.08 &     0.37 &    14.63  \\
     502 &   03 44 38.84 &    32 14 47.7 &    1.86 &    19.14 &     0.89 &     1.12 &     0.45 &    14.67  \\
     509 &   03 45 06.97 &    32 14 03.9 &    1.36 &    18.37 &       \nodata &     0.98 &     0.37 &    15.12  \\
     546 &   03 44 27.83 &    32 08 00.4 &    1.41 &    20.41 &     0.83 &     0.93 &     0.50 &    16.00  \\
     574 &   03 44 52.36 &    32 11 27.4 &    1.27 &    19.60 &     0.71 &     0.75 &     0.45 &    16.14  \\
     581 &   03 44 41.24 &    32 12 16.4 &      \nodata &    21.93 &     0.20 &     1.61 &     0.70 &    16.39  \\
     584 &   03 44 43.82 &    32 11 54.1 &      \nodata &       \nodata &       \nodata &     2.20 &     0.92 &    15.90  \\
     588 &   03 44 24.57 &    32 11 40.6 &      \nodata &    20.64 &     0.95 &     1.42 &     0.57 &    15.56  \\
     594 &   03 44 38.58 &    32 11 04.8 &      \nodata &    21.90 &     1.23 &     1.70 &     0.61 &    16.08  \\
     745 &   03 44 24.29 &    32 06 11.8 &      \nodata &    20.48 &     0.97 &     1.40 &     0.39 &    15.77  \\
     901 &   03 44 50.47 &    32 12 09.5 &    1.30 &    20.47 &     0.68 &     1.07 &     0.60 &    16.25  \\
     917 &   03 44 48.36 &    32 12 26.4 &      \nodata &    20.72 &     1.04 &     1.34 &     0.58 &    15.66  \\
     943 &   03 44 16.36 &    32 12 53.9 &    1.23 &    18.72 &     0.47 &     0.72 &     0.28 &    16.09  \\
     951 &   03 44 10.86 &    32 13 00.6 &    1.15 &    18.13 &     0.52 &     0.86 &     0.17 &    15.44  \\
     959 &   03 44 09.80 &    32 13 07.1 &    1.44 &    19.75 &     0.60 &     0.85 &     0.49 &    16.34  \\
     998 &   03 44 53.49 &    32 13 56.9 &    1.60 &    20.07 &     0.75 &     1.00 &     0.38 &    16.25  \\
    1007 &   03 45 02.81 &    32 14 07.6 &    2.47 &    19.80 &     0.92 &     1.10 &     0.58 &    15.38  \\
    1020 &   03 44 36.31 &    32 14 19.9 &      \nodata &    20.87 &     1.05 &     1.54 &     0.72 &    14.98  \\
    1021 &   03 44 59.07 &    32 14 22.8 &    1.47 &    18.67 &     0.72 &     0.95 &     0.37 &    15.02  \\
    1062 &   03 44 48.14 &    32 15 22.4 &      \nodata &    19.47 &     0.80 &     1.14 &     0.44 &    14.87  \\
    1110 &   03 45 00.78 &    32 16 35.7 &    1.71 &    20.14 &     0.74 &     1.08 &     0.41 &    16.07  \\
    1139 &   03 44 59.68 &    32 17 30.3 &    1.84 &    19.01 &     0.78 &     1.23 &     0.42 &    14.58  \\
    1142 &   03 44 55.39 &    32 17 34.9 &    1.65 &    19.28 &     0.75 &     1.04 &     0.37 &    15.53  \\
    1167 &   03 44 32.68 &    32 18 06.1 &    1.30 &    19.99 &       \nodata &     1.16 &     0.51 &    15.99  \\
    1173 &   03 44 53.81 &    32 18 12.1 &    1.60 &    19.23 &     0.70 &     0.87 &     0.34 &    15.87  \\
    1194 &   03 45 02.04 &    32 18 36.3 &    2.10 &    18.36 &     0.86 &     1.25 &     0.46 &    13.93  \\
    1555 &   03 44 23.09 &    32 15 12.2 &      \nodata &    19.86 &     0.73 &     1.27 &     0.29 &    16.21  \\
    1729 &   03 44 25.41 &    32 13 56.5 &    1.38 &    19.05 &     0.69 &     0.83 &     0.34 &    15.76  \\
    1735 &   03 44 08.33 &    32 13 57.4 &    1.18 &    19.54 &     0.59 &     0.64 &     0.01 &    17.13  \\
    1742 &   03 44 21.65 &    32 13 47.8 &    1.31 &    17.65 &     0.59 &     0.85 &     0.26 &    14.75  \\
    3076 &   03 44 56.39 &    32 13 35.8 &      \nodata &    21.32 &     0.98 &     0.79 &     0.82 &    16.66  \\
    3104 &   03 45 01.75 &    32 14 34.4 &      \nodata &    20.04 &     0.78 &     0.69 &     0.62 &    15.00  \\
    3117 &   03 44 49.41 &    32 15 04.8 &      \nodata &    22.16 &     1.04 &     1.62 &     0.73 &    16.21  \\
    3215 &   03 44 46.55 &    32 13 50.5 &      \nodata &    22.04 &     1.03 &     1.52 &     0.61 &    16.36  \\
    4053 &   03 44 14.89 &    32 06 12.2 &      \nodata &    21.58 &     1.07 &     1.36 &     0.77 &    15.70  \\
   22253 &   03 44 23.73 &    32 17 16.2 &      \nodata &    18.28 &     0.57 &     0.02 &     0.09 &    15.55  \\
\enddata
\tablenotetext{a}{From the 2MASS Point Source Catalog
for 509, from \citet{luh99} for 584 and 1167, and from \citet{luh03b}
for the remaining objects.}
\tablenotetext{b}{From \citet{her98} for 222 and 546 and 
from \citet{luh99} for the remaining objects.}
\tablenotetext{c}{From \citet{luh99} for 411, 509, and 1167
from \citet{luh03b} for the remaining objects.}
\tablenotetext{d}{\citet{luh03b}.}
\tablenotetext{e}{\citet{mue03}.}
\end{deluxetable}

\end{document}